\documentclass[twocolumn]{revtex4-2}
\usepackage{amssymb}
\usepackage{xcolor}
\usepackage{graphicx}

\begin{document}
\title{High magnetic field induced crossover from the Kondo to Fermi liquid behavior in 1$T$-VTe$_{2}$ single crystals}
\author{Xiaxin Ding$^{1,2}$}
\email{xiaxin.ding@inl.gov}
\author{Jie Xing$^2$}
\author{Gang Li$^3$}
\author{Luis Balicas$^3$}
\author{Krzysztof Gofryk$^1$}
\author{Hai-Hu Wen$^2$}

\affiliation{$^1$Idaho National Laboratory, Idaho Falls, Idaho, 83402, USA.}
\affiliation{$^2$Center for Superconducting Physics and Materials, National Laboratory of Solid State Microstructures and Department of Physics, Nanjing University, Nanjing 210093, China}
\affiliation{$^3$National High Magnetic Field Laboratory, Florida State University, Tallahassee-FL 32310, USA}

\begin{abstract}
The magnetic and magnetotransport properties of metallic 1$T$-VTe$_{2}$ single crystals were investigated at temperatures from 1.3 to 300 K and in magnetic fields up to 35 T. Upon applying a high magnetic field, it is found that the electrical resistivity displays a crossover from the logarithmic divergence of the single-impurity Kondo effect to the Fermi liquid behavior at low temperatures. The Brillouin scale of the negative magnetoresistivity above the Kondo temperature $T_{\rm{K}}$ = 12 K indicates that the Kondo features originate from intercalated V ions, with $S$ = 1/2. Both magnetic susceptibility and Hall effect show an anomaly around $T_{\rm{K}}$. By using the modified Hamann expression we successfully describe the temperature-dependent resistivity under various magnetic fields, which shows the characteristic peak below $T_{\rm{K}}$ due to the splitting of the Kondo resonance.

\end{abstract}

\maketitle

\section{INTRODUCTION}

Layered transition metal dichalcogenides (TMDCs) have attracted enormous attention because they exhibit a rich variety of physical properties and a striking potential for applications~\cite{wilson1969transition,wang2012electronics,chhowalla2013chemistry}. The wide range of transport properties of bulk TMDCs varies from conventional superconductivity~\cite{morris1972superconductivity} via charge density wave (CDW)~\cite{CDW} to exceedingly large positive magnetoresistivity (MR)~\cite{ali2014large} at low temperatures. Furthermore, layered TMDCs are easily intercalated with metallic ions due to their layered structures~\cite{WHITTINGHAM197841}. For example, Cu intercalated TiSe$_2$ exhibits a CDW to superconducting transition upon doping~\cite{Morosan2006}. When magnetic 3$d$ transition metals (Co, Ni and Fe) are intercalated into TiSe$_2$, the Kondo effect is induced in the dilute limit~\cite{PhysRevB.82.224416}. Similarly, the Kondo effect has been reported in bulk VSe$_2$, probably due to the interlayer V ions~\cite{VSe2}. In addition, TMDCs are ideal for studying the interplay between electronic properties and crystalline dimensionality. With the development of exfoliation and epitaxial techniques, monolayered TMDCs could be successfully produced demonstrating distinct properties from their bulk counterparts. In contrast to paramagnetism in bulk VSe$_2$, ferromagnetism is observed in monolayer VSe$_2$ at room temperature~\cite{Bonilla2018}. In monolayer VTe$_2$, which is isostructural to VSe$_2$, the suppression of a CDW transition is a matter of debate~\cite{Coelho,ARPES,ARPES2,PhysRevB.101.205105}. In bulk VTe$_2$, a CDW transition occurs upon cooling across 480 K, which is accompanied by a structural phase transition from the high-temperature trigonal phase (1$T$ structure) to the low-temperature monoclinic phase~\cite{ohtani1981phase,BRONSEMA1984415,Kamitani2015,Mitsuishi2020}. Using the vapor transport method, the single crystalline VTe$_2$ has a monoclinic structure corresponding to the low-temperature polymorph~\cite{mono}. By contrast, high-temperature polymorph 1$T$-VTe$_2$ single crystals can be successfully grown via the molten-salt method. Furthermore, this reveals characteristics of the Kondo effect at low temperatures~\cite{dingthesis}, which was later found in 1$T$-VTe$_2$ nanoplates~\cite{VTe2Nano}.

In this manuscript, we study the evolution of the Kondo effect under high magnetic fields in 1$T$-VTe$_2$ single crystals. The Kondo effect, discovered in 1930 by Meissner and Voigt~\cite{MandV} and then explained by Kondo in 1964~\cite{Kondo1964}, has recently generated recently renewed interest due to progress in nanotechnology~\cite{Zhao1542,Calvo2009}. In condensed matter physics, it also provides clues to allow understanding of the electronic properties of various strange metals, such as superconductors and heavy-fermion materials. The spin-exchange process between localized impurities and itinerant electrons of the metallic host generates a new state at the Fermi level, known as the Kondo resonance or Abrikosov-Suhl resonance peak~\cite{Abrikosov1970}. The first hint of this new state manifests itself as an anomalous upturn in the resistivity below the Kondo temperature $T_{\rm{K}}$, which is the energy scale dependent on spin-exchange coupling and limiting the validity of the Kondo effect. Specifically, the temperature-dependent resistivity $\rho(T)$ of the Kondo system shows the approximate logarithmic increase with decreasing temperature in a small range below $T_{\rm{K}}$ and $T^2$ Fermi liquid behavior for $T \ll T_{\rm{K}}$. Moreover, the splitting of the Kondo resonance under the applied magnetic field correlates with a characteristic peak in the temperature and magnetic field dependence of the resistivity $\rho(T,B)$. These transport properties for a single Kondo impurity can be calculated by a nonperturbative approach, such as the numerical renormalization group (NRG) method~\cite{RevModPhys.47.773,Costi2000kondo}. However, despite all of these advances, the lack of a systematic understanding remains on the low-temperature behavior of the electrical resistivity, especially in the presence of high magnetic fields~\cite{felsch1973magnetoresistivity}.

Here, we systematically study the transport and magnetic properties of 1$T$-VTe$_2$ single crystals under high magnetic fields. The temperature and magnetic field dependence of the magnetization, resistivity, and Hall effect reveal Kondo effect characteristics with $T_{\rm{K}}$ = 12 K. Both magnetic susceptibility and MR measurements determined that the local magnetic moments arise from the intercalated V ions with $S$ = 1/2. The characteristic peaks in $\rho(T,B)$ are observed below $T_{\rm{K}}$ and can be analyzed by a modified Hamann expression. By applying a magnetic field up to 35 T, the Kondo effect is gradually suppressed, and the Fermi liquid behavior emerges at low temperatures.

\section{EXPERIMENTAL DETAILS}

1$T$-VTe$_2$ single crystals were grown by the flux method using KCl as the flux. First, we prepared the polycrystalline samples by a solid-state reaction method, with V powders (purity 99.5\%, Alfa Aesar) and Te grains (purity 99.5\%, Alfa Aesar) in the ratio of 1 : 2. The mixture was compressed into a pellet, then loaded into an alumina crucible and sealed in an evacuated quartz tube, subsequently heated up to 750 $^{\circ}$C for 20 h. Second, powders with a molar ratio of KCl : VTe$_2$ = 4 : 1 were heated up to 950 $^{\circ}$C for 2 days, followed by cooling down to 800 $^{\circ}$C at a rate of 1 $^{\circ}$C/h. Finally, we obtained single crystals with lateral sizes of 2-4 mm and thickness of about 10-50 $\mu$m by dissolving the flux in deionized water. The samples are stable in water and display silver color. All the weighing, mixing, grinding and pressing procedures were finished in a glovebox under Ar atmosphere. X-ray diffraction (XRD) measurements were performed using a Bruker D8 Advanced diffractometer with Cu K$_\alpha$ radiation. The energy dispersive X-ray (EDX) spectrum measurements were performed on a scanning electron microscope under an accelerating voltage of 20 kV (Hitachi Co., Ltd.). The magnetization measurements were carried out using a SQUID-VSM-7T Quantum Design device. The small enhancement, around 50 K in the temperature dependence of the magnetic susceptibility at 1 T, is induced by an antiferromagnetic transition due to a small amount of solid oxygen in the measurement chamber. Electrical transport measurements were done in a Quantum Design PPMS-16T instrument using a standard four-probe method, with the electrical current applied along the plane of samples. The high field MR was measured at the National High Magnetic Field Laboratory in Tallahassee.

\section{RESULTS and DISCUSSION}

\subsection{X-ray diffraction}

\begin{figure}[htbp]
\includegraphics[width=0.5\textwidth]{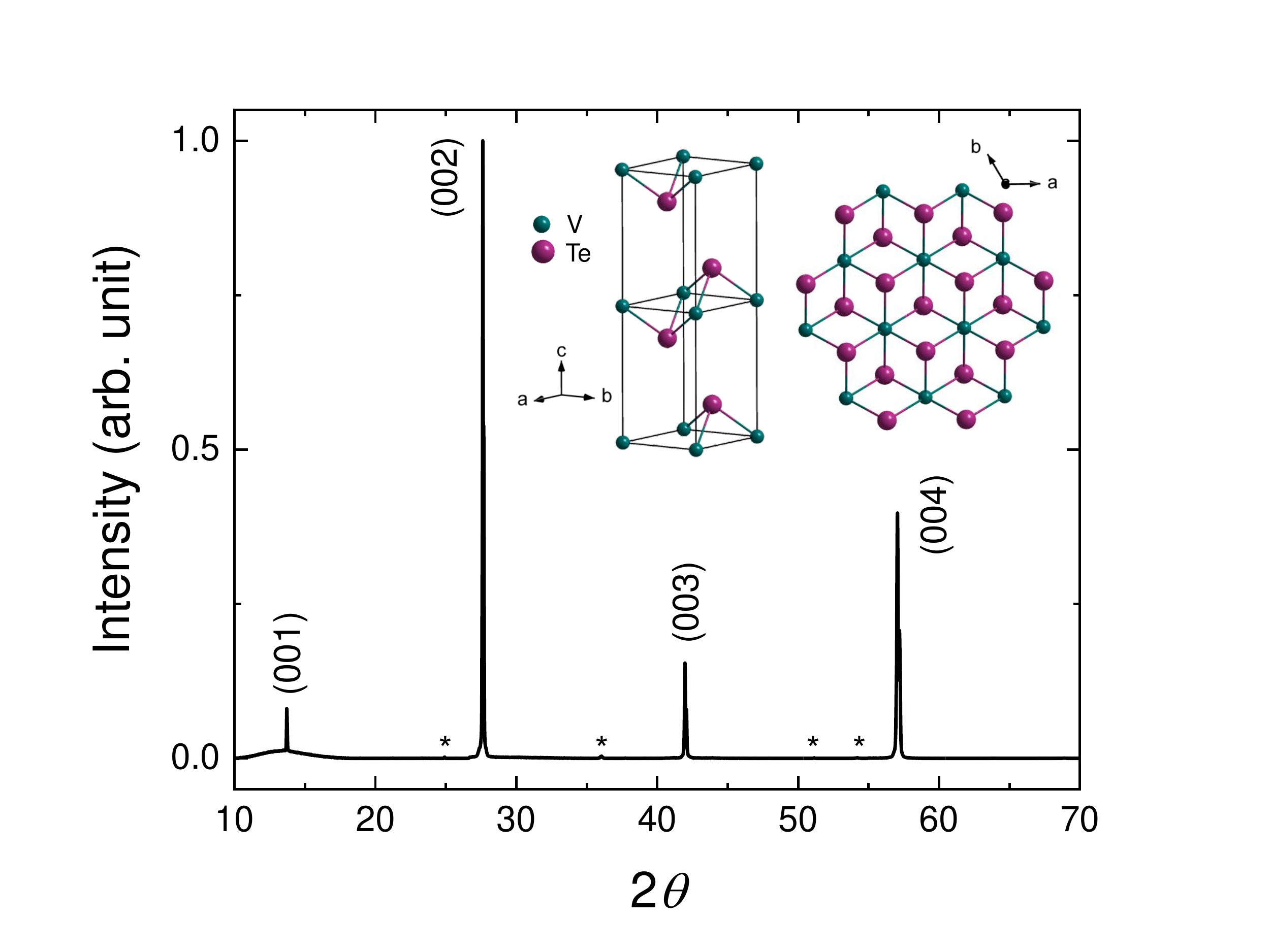}
\caption {(color online) X-ray diffraction patterns of 1$T$-VTe$_2$ single crystals. Asterisks mark peaks that do not belong to the trigonal phase. Inset: Schematic atomic structure of 1$T$-VTe$_2$. Olive and purple circles represent V and Te atoms, respectively.}\label{X}
\end{figure}

The XRD pattern of the as-grown 1$T$-VTe$_2$ single crystals at room temperature is shown in Fig.~\ref{X}. Similar to its sister compounds~\cite{VANBRUGGEN1976251,BAYARD1976325,MURPHY1979339,Kamitani2015}, our VTe$_2$ single crystals display a trigonal CdI$_2$-type structure (1$T$ phase) with V atoms (olive circles) located at the center of the octahedra formed by Te atoms (purple circles), as shown in the inset of Fig.~\ref{X}. The V-Te layers are stacked along the $c$-axis by the van der Waals-like forces. As suggested by the transport and magnetic properties shown below, there is no additional structural transition down to 1.3 K. Sharp (00$l$) Bragg peaks can be observed and yield a lattice constant $c$ = 6.456 $\rm\mathring{A}$, which is similar to those of 1$T$-V$_{0.7}$Ti$_{0.3}$Te$_2$ single crystals~\cite{Kamitani2015} and 1$T$-VTe$_2$ nanoplates~\cite{VTe2Nano}. For single crystals made by the vapor transport method, the resulting CDW state leads to a monoclinic structure with $c$ = 9.069 $\rm\mathring{A}$~\cite{BRONSEMA1984415,mono}. As marked by asterisks in Fig.~\ref{X}, a very small amount of unknown impurity phases has been identified. It corresponds to less than 2\% of the sample volume and does not impact our studies or conclusions presented in the paper. The EDX analysis performed at different locations on the crystal surface of 1$T$-VTe$_2$ yields a V:Te composition close to (1.01$\pm$0.03) : 2.

\subsection{Resistivity and magnetic susceptibility}

\begin{figure}[htbp]
\includegraphics[width=0.5\textwidth]{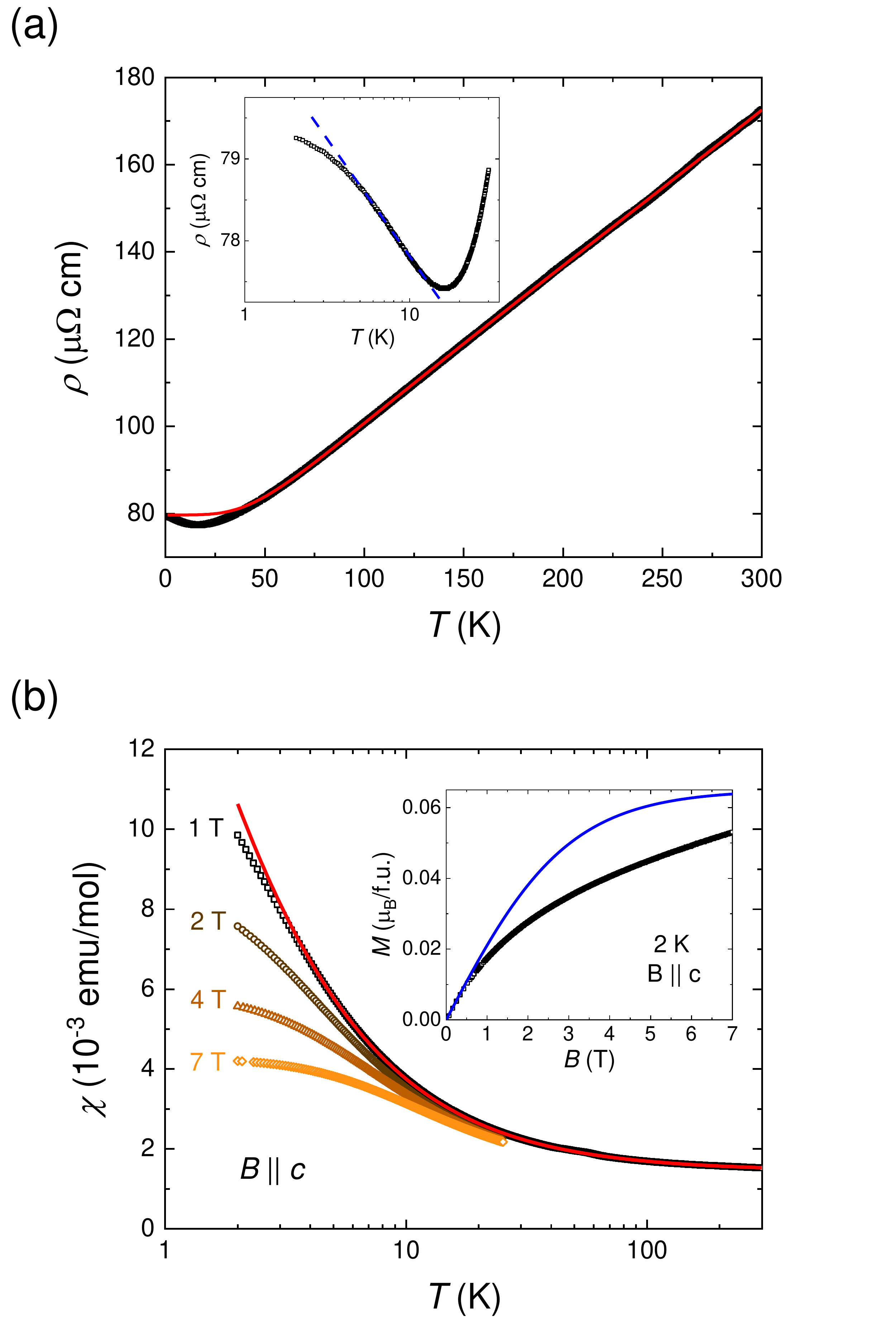}
\caption {(color online) (a) Temperature dependence of the in-plane electrical resistivity of VTe$_2$. Red curve is the fitting result of Eq. 1. Inset: An enlarged view of low temperatures in semi-log plots. Blue dashed line is a guide for the eye and shows a -$lnT$ dependence. (b) Semi-log plots of the temperature dependence of the magnetic susceptibility of VTe$_2$, measured under several magnetic fields applied along the $c$-axis. Red line is the Curie-Weiss fit in the temperature range from 5 - 300 K at 1 T. Inset: Magnetic field dependence of the magnetization measured at 2 K. Blue line is calculated $M(B)$ using the Brillouin function Eq. (4).}\label{RT}
\end{figure}

The temperature dependence of the electrical resistivity $\rho(T)$ is illustrated in Fig.~\ref{RT}(a). At 300 K, the value of $\rho$ is about 172.5 $\mu\Omega$ cm, close to that of V$_{0.7}$Ti$_{0.3}$Te$_2$ single crystals~\cite{Kamitani2015} while about an order of magnitude smaller than that of VTe$_2$ nanoplates~\cite{VTe2Nano}. In contrast to the CDW transition observed in $\rho(T)$ of V$_{1-x}$Ti$_x$Te$_2$ single crystals~\cite{Kamitani2015}, no signature for a structural transition is found down to 2 K in VTe$_2$ single crystals. The $\rho(T)$ decreases from room temperature to a minimum value near 16 K. Rather than saturating at low temperatures, as expected for simple metals, the resistivity displays an increase with decreasing temperature below the minimum. The inset of Fig.~\ref{RT}(a) shows the low-temperature behavior on semi-log plots. The logarithmic increase below 16 K is characteristic of Kondo systems and it is due to a contribution from the conduction electron-magnetic impurity interaction~\cite{Kondo1964}. Moreover, there is a divergence from the logarithmic increase below 5 K, mainly due to the spin-compensated state or the Ruderman-Kittel-Kasuya-Yosida (RKKY) interactions between magnetic impurities~\cite{Wassermann,Liang}. The residual-resistivity ratio, defined as $\rho$(300 K)/$\rho$(16 K), is relatively small and estimated to be only 2.2 in VTe$_2$ single crystals. This is consistent with the presence of small amounts of impurities in the Kondo system. Therefore, the measured resistivity is the sum of the typical electron-phonon and electron-electron interactions, and the spin scattering of the conduction electrons by the magnetic impurity characteristic of the Kondo effect. First, we focus on the high-temperature behavior of the resistivity and consider the Kondo term $\rho_{\rm{K}}$ as a constant. As shown by the red line in Fig.~\ref{RT}(a), we fit $\rho(T)$ of 1$T$-VTe$_2$ in the temperature range 50 - 300 K to the formula
\begin{equation}
\rho(T) = \rho_0 + \rho_{\rm{K}} + aT^2 + \rho_{\rm{ph}}(T),  
\end{equation}
where $\rho_0 + \rho_{\rm{K}}$ = 79.68 $\mu\Omega$ cm is the sum of the residual resistivity and the Kondo contribution, the $aT^2$ term represents the electron-electron interaction, and $\rho_{\rm{ph}}(T)$ arises from electron-phonon interactions and is expressed by the Bloch-Gr$\rm\ddot{u}$eisen formula
\begin{equation}
\rho_{\rm{ph}}(T) = \alpha\left(\frac{T}{\Theta_{\rm{R}}}\right)^5\int_{0}^{\frac{\Theta_{\rm{R}}}{T}} \frac{x^5}{(e^x-1)(1-e^x)}dx.
\end{equation}
In the model, $\alpha$ is a constant proportional to the electron-phonon coupling and $\Theta_R$ is the Debye temperature. The value $\Theta_{\rm{R}}$ = 274 K obtained from our electrical resistivity analysis is close to $\Theta_{\rm{D}}$ = 267 $\pm$ 20 K, as obtained from heat capacity measurement of VTe$_2$ polycrystals~\cite{DebyeTemp}. This approach fails to describe the low-temperature behavior of VTe$_2$ which can be well explained by including the Kondo effect. The analysis of the low-temperature part of $\rho(T)$ will be discussed below.

Figure~\ref{RT}(b) shows the temperature dependence of the magnetic susceptibility $\chi(T)$, measured under several values of the magnetic field along the $c$-axis. The susceptibility increases monotonically with decreasing temperature under 1 T. No magnetic order is observed down to 2 K. As shown by the red curve, the $\chi(T)$ data from 5 - 300 K could be well fitted to a modified Curie-Weiss formula 
\begin{equation}
\chi(T) = \chi_0 - fT + \frac{C}{T-\theta},
\end{equation}
where $\chi_0 = 1.45 \times 10^{-3}$ emu/mol is the Pauli susceptibility of conduction electrons in the host VTe$_2$. In general, the Pauli susceptibility of a metallic sample is temperature independent. However, in our case, a phenomenological fitting term $-fT$ with  $f = 2.79 \times 10^{-7}$ emu/(mol K) needs to be added. The temperature-dependent Pauli susceptibility might be due to the change of relative sizes of the Fermi surface and Brillouin zones, where the crystal contract anisotopically with decreasing temperature. For instance, such behavior has been observed in $\alpha$-U single crystals~\cite{alpha}. Similar linear temperature dependence for $\chi$ has been observed in the sister compound NiTe$_2$~\cite{NiTe2}, in which spin-polarized topological surface states were observed~\cite{PhysRevB.100.195134}. The third term is the Curie-Weiss susceptibility coming from Kondo impurities. As shown in the main frame of Fig.~\ref{RT}(b), the deviation of $\chi(T)$ below 4 K from the fitting is further indication of Kondo behavior. It demonstrates the Kondo screening of the magnetic moment on the impurity from the surrounding cloud of negatively polarized conduction electrons, combined with the possible onset of short range order, mediated by an RKKY interaction between the intercalated V impurities. As the temperature decreases, the magnetic moment of the Kondo impurity crossover from localized behavior at high temperatures, described by the Curie-Weiss law, to a fully compensated moment at low temperatures, where temperature-independent Pauli susceptibility is formed~\cite{Park2011field}. We obtain the values of the Curie constant $C$ = 0.0246 emu K/mol and the Curie-Weiss temperature $\theta$ = -0.676 K. According to the EDX analysis, there are no other magnetic elements in the sample. Furthermore, layered compounds are easily intercalated with metallic guests due to their low-dimensional structures~\cite{WHITTINGHAM197841}. Thus, we assume that the intercalated V ions are indeed the magnetic impurities in the VTe$_2$ sample. Taking the spin of the localized V$^{4+}$ ($S$ = 1/2) into account, the theoretical magnetic moment of each scattering center is $\mu_{\rm{V}} = g\sqrt{S(S+1)}\rm{\mu_B} = 1.73~\rm{\mu_B}$, where $g$ = 2 assuming quenched orbital contribution. Using the effective moment of the sample $\mu_{\rm{eff}} \approx \sqrt{8C}$ = 0.443~$\rm{\mu_B}$/f.u., we estimate that the molar fraction of intercalated V ions is $N_1 = \mu^2_{\rm{eff}}/\mu^2_{\rm{V}}$ = 0.066. As shown in Fig.~\ref{RT}(b), similar to VSe$_2$~\cite{VSe2}, the deviation of the susceptibility with respect to the Curie-Weiss law is more significant at higher magnetic fields. Moreover, the decrease of the saturated susceptibility indicates a reduction in the compensated moment of the impurity by the magnetic field. The inset of Fig.~\ref{RT}(b) shows the magnetic field dependence of the magnetization at 2 K. The induced magnetization is only 0.05 $\rm{\mu_B}$ at 2 K and 7 T. By taking account the $N_1$ value obtained above, the magnetization could be calculated using the Brillouin function $B_J(x)$ with $J$ = 1/2 as
\begin{equation}
M_{\rm{cal}} = N_1g\rm{\mu_B}B_J\left(\frac{g\rm{\mu_B}JB}{k_{\rm{B}}T}\right).
\end{equation}
As shown by the blue line in Fig.~\ref{RT}(b), the $M_{\rm{cal}}(B)$ curve is compared to the measured data. The deviation above 0.3 T is mainly related to the splitting of the Kondo resonance under magnetic fields.

\subsection{Magnetoresistivity}

\begin{figure}[htbp]
\includegraphics[width= 0.5\textwidth]{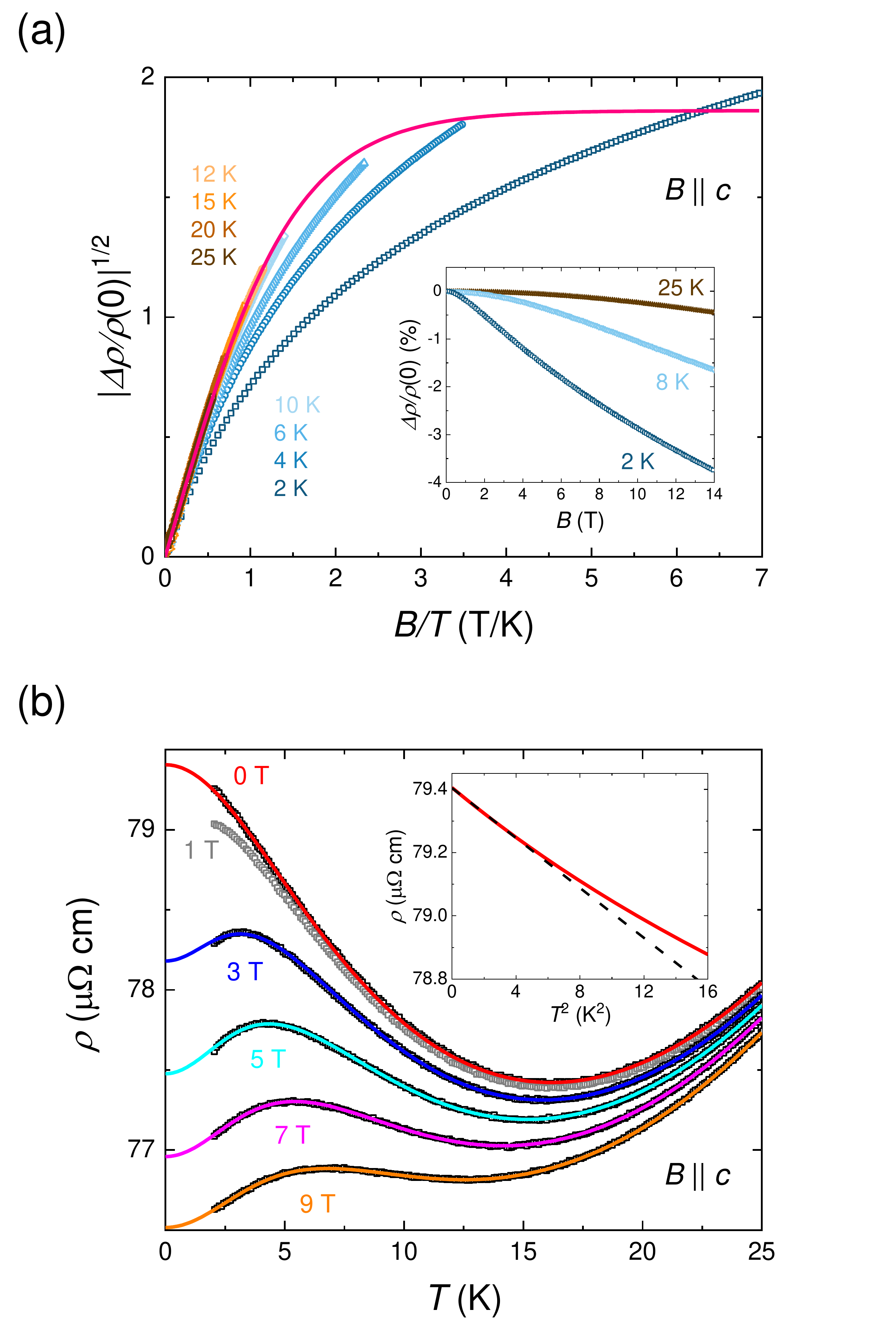}
\caption {(color online) (a) Square root of $|\Delta\rho/\rho(0)|$ as a function of $B/T$ measured at various temperatures. Pink line is the Brillouin function with a scaled magnitude. Inset: Magnetic field dependence of the MR at different temperatures. (b) Temperature dependence of the resistivity under different magnetic fields. Solid lines are fits to the modified Hamann expression Eq. (6) and (8). Inset: An enlarged view of the fitting curve under zero magnetic field is shown by the red line. Black dashed line is a guide for the eye and shows a -$T^2$ dependence.}\label{MR}
\end{figure}

To further evaluate the magnetic Kondo impurity, we systematically measured the magnetic field dependence of MR, $\Delta\rho/\rho(0) = [\rho(B)-\rho(0)]/\rho(0) \times 100\%$ along the $c$-axis at different temperatures. The results are presented in the inset of Fig.~\ref{MR}(a). At 25 K, the negative MR($B$) shows a convex shape in the entire field range. In contrast, the nonlinear MR at 2 K slightly bends upwards under high fields and reaches -3.7\% at 14 T. The negative MR has been derived by Yosida from the first-order perturbation of the $s$-$d$ exchange interaction~\cite{yosida1957anomalous}. If magnetic moments existed, the magnitude of the negative MR should be proportional to the square of the impurity magnetization, for $T > T_K$ and in the low $B/T$ limit~\cite{PhysRev.170.552}. Accordingly, we plot $|\Delta\rho/\rho(0)|$ as a function of $B/T$ for different temperatures in Fig. \ref{MR}(a). It is clear that the negative MR above 12 K collapses and can be scaled by the Brillouin function ($J$ = 1/2)
\begin{equation}
\left|\frac{\Delta\rho}{\rho(0)}\right|^\frac{1}{2} = \lambda B_J\left(\frac{g\rm{\mu_B}JB}{k_{\rm{B}}T}\right),
\end{equation}
where $\lambda$ is a scaling constant. These results strongly support the scenario of the localized V impurity with $S$ = 1/2. The gradual deviation from the scaling function below 12 K indicates the screening conduction electrons, which is a scale determined by many-body interactions. Thus, we obtain the Kondo temperature $T_{\rm{K}}$ = 12 K, where the Kondo resonance is gradually formed.

In order to analyze the Kondo effect in VTe$_2$ in greater detail, we focus on the low-temperature behavior of the resistivity in this system. Fig.~\ref{MR}(b) shows the temperature dependence of resistivity from 2 to 25 K and under different magnetic fields applied along the $c$-axis. Since the phonon contribution is small at low temperatures and doesn't change under magnetic fields, we analyze the low-temperature data by taking into account the Fermi liquid behavior and the contribution from the Kondo scattering only. The Hamann expression, yielding a new solution for the $s$-$d$ exchange model, is mostly used for analyzing the Kondo effect in the low-temperature electrical resistivity. It was reduced from Nagaoka's self-consistent equations to a single nonlinear integral equation~\cite{hamann1967new}. However, its applicability is limited to temperatures $T \ge T_{\rm{K}}$. For $T < T_{\rm{K}}$, the Hamann expression gives low values for the impurity spin $S$ due to inadequacies in the Nagaoka's approximation~\cite{Nagaoka}. To account for this issue, we replace the variable $T$ with $T_{\rm{eff}} = \sqrt{T^2 + T_{\rm{W}}^2}$, where $k_{\rm{B}}T_{\rm{W}}$ is the effective RKKY interactions~\cite{FISCHER,Wassermann}. As demonstrated by the red line, $\rho(T)$ under zero field can be analyzed by the formula
\begin{equation}
\rho(T) = \rho_0 + aT^2 + \rho_{\rm{H}}(T_{\rm{eff}}),
\end{equation}
where $\rho_0$ is the residual resistivity, $\rho_{\rm{H}}(T_{\rm{eff}})$ is the modified Hamann expression~\cite{hamann1967new, FISCHER, Wassermann}, which leads to the crossover region and to saturation at low temperatures, and is given by
\begin{equation}
\rho_{\rm{H}}(T_{\rm{eff}}) = \rho_{\rm{K}}\left[1-\frac{ln(T_{\rm{eff}}/T_{\rm{K}})}{\sqrt{ln^2(T_{\rm{eff}}/T_{\rm{K}})+\pi^2 S(S+1)}}\right],
\end{equation}
where $\rho_{K}$ is a temperature-independent constant, $T_{\rm{W}} = 5.18$ K is related to the average RKKY interaction strength between V impurities and its value is close to the deviation temperature of the logarithmic increase. According to the above MR analysis, the spin of the magnetic impurities $S$ and the Kondo temperature $T_{\rm{K}}$ are fixed to 1/2 and 12 K, respectively. As shown in the inset of Fig. 3(b), in the Kondo regime at $T \ll T_{\rm{K}}$, the red fitting curve exhibits the -$T^2$ dependence below 2 K, characteristics of single-impurity Kondo effect and Nagaoka theory~\cite{Wassermann,Nagaoka,Gauzzi_2019,Nozieres}. Thus, the modified Hamann expression gives a phenomenologically satisfactory description of the logarithmic increase with decreasing temperature in a small range below $T_{\rm{K}}$ and the -$T^2$ behavior at $T \ll T_{\rm{K}}$ in $\rho(T)$~\cite{Costi_1994}.

Because the system shows negative MR, it is expected that the upturn of $\rho(T,B)$ is suppressed with increasing magnetic fields. Moreover, the splitting of the Kondo resonance under applied magnetic fields correlates with the electrical transport property, in which $\rho(T,B)$ shows an extra increase with decreasing temperature below $T_{\rm{K}}$. As shown in Fig.~\ref{MR}(b), a finite temperature peak is observed when the applied magnetic field is larger than 3 T. With increasing magnetic field, the peak becomes broader and shifts to higher temperatures. Motivated by previous theoretical calculations~\cite{PhysRev.170.552} and experimental analysis of Ce$_x$La$_{1-x}$Al$_2$ alloys~\cite{felsch1973magnetoresistivity}, an impurity spin polarization function [1-$L^2(x)$] is added to the modified Hamann expression in Eq.~(7) for the analysis of $\rho(T,B)$. Not only magnetic impurities interact with the conduction electrons below $T_{\rm{K}}$, but also the relaxation time describing the Kondo resonance is different for up spins and down spins when a large enough magnetic field is applied~\cite{yosida1957anomalous}. Therefore, we use the Langevin function $L(x)$ instead of the Brillouin function $B_J(x)$ for the analysis of $\rho(T,B)$~\cite{felsch1973magnetoresistivity}. Furthermore, the variable $B/T$ of the Langevin function is replaced by $B/T_{\rm{eff}}$
\begin{eqnarray}
\nonumber \rho_{\rm{H}}(T_{\rm{eff}}, B) &=& \rho_{\rm{K}}\left[1-\frac{ln(T_{\rm{eff}}/T_{\rm{K}})}{\sqrt{ln^2(T_{\rm{eff}}/T_{\rm{K}})+\pi^2 S(S+1)}}\right]\\
 &&\cdot\left\{1-L^2\left[\frac{\mu B}{k_{\rm{B}}T_{\rm{eff}}}\right]\right\},
\end{eqnarray}
where $\mu$ is the effective magnetic moment of the vanadium impurity. The values for $T_{\rm{K}}$ and $S$, as fixed in the fitting of the zero-field resistivity, are left unchanged. As shown in Fig.~\ref{MR}(b), the change of MR at 1 T is relatively small. This indicates that magnetic field might not be strong enough to split the Kondo resonance at finite temperatures~\cite{Costi2000kondo}. Also, for a reliable analysis of the result by Eq.(8) at low magnetic fields, measurements at lower temperatures are required (below 2 K). All fitting parameters are listed in Table I. The term $\rho_0 + \rho_{\rm{K}}$ gradually decreases with an increasing magnetic field, which is consistent with the negative MR. The reduction of $\mu$ indicates that the impurity gradually loses its magnetic character under magnetic fields. This is also reflected in the magnetic susceptibility measurements in Fig.~\ref{RT}(b). The 3$d$ vanadium impurity is unique, because its magnetic moment varies from 0 in bulk V and clusters of V atoms, via 3/5 $\rm{\mu_B}$ in the atomic model, to a maximum of 3 $\rm{\mu_B}$ in the $d$ resonance model~\cite{PhysRevLett.88.167202}. In VSe$_2$, previous work suggests that the intercalated V ion produces a net paramagnetic moment of 2.5 $\rm{\mu_B}$~\cite{VSe2, PhysRevB.19.3420}. A related point to consider is that a magnetic moment as large as 6.5 $\rm{\mu_B}$ was reported for V impurities in thin films of Na host, indicating a polarization of the host~\cite{PhysRevLett.88.167202}. In the last section, we used the magnetic moment of quenched V$^{4+}$ ions to estimate the molar fraction of V impurities. Here, using $\mu (\rm{3~T}) = 2.436~\rm{\mu_B}$ from the Langevin function fit, the molar fraction is estimated to be $N_2 = \mu^2_{\rm{eff}}/\mu^2$ = 0.033, which is closer to the EDX result for the deviation from stoichiometry $N$ = 0.01$\pm$0.03.

\begin{figure}[htbp]
\includegraphics[width=0.5\textwidth]{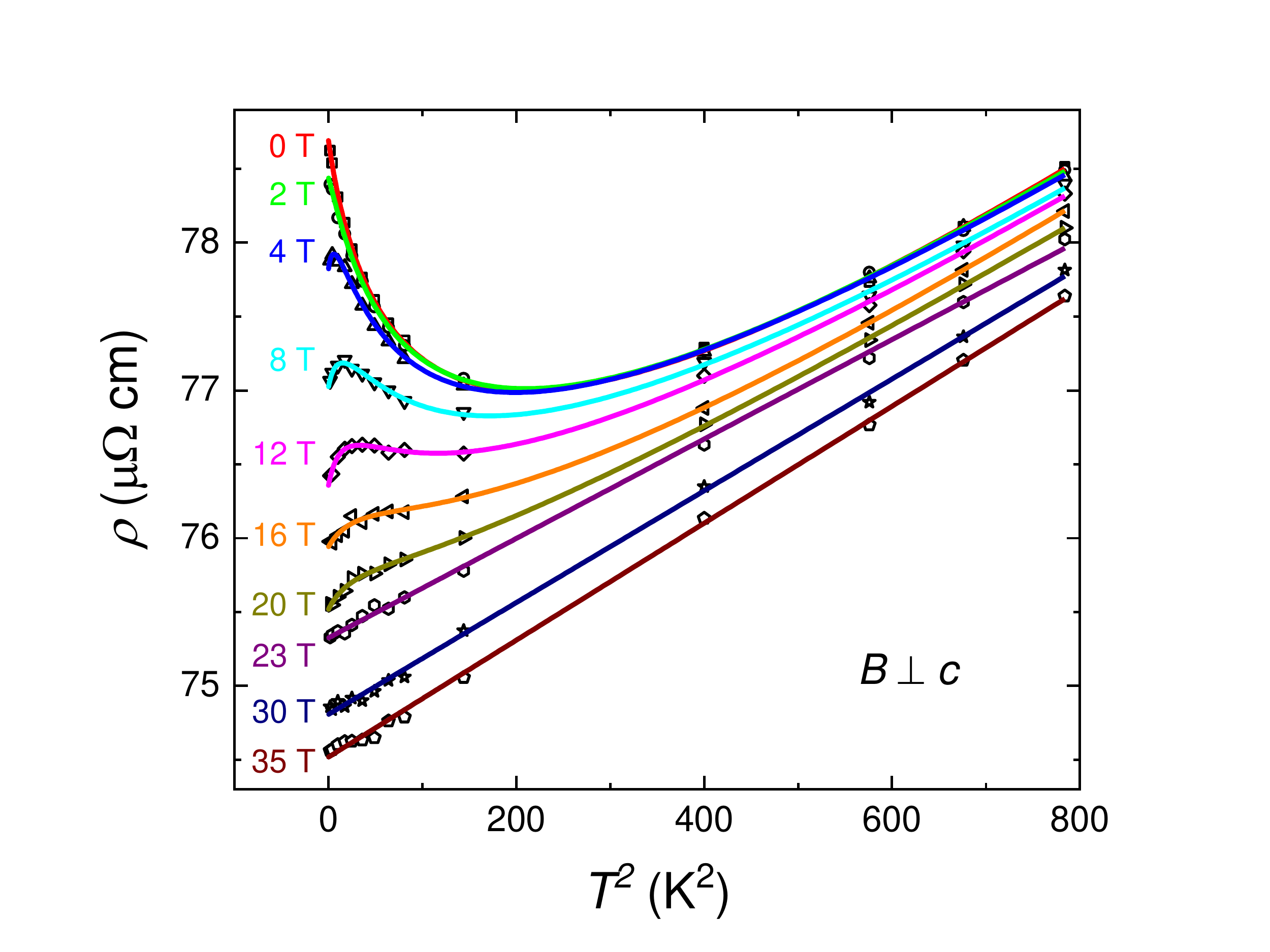}
\caption {(color online) Resistivity as a function of $T^2$ measured under different magnetic fields applied perpendicularly to the $c$-axis. Discrete data points were determined from the measurements by sweeping the magnetic field at a fixed temperature. Solid lines are fits at different magnetic fields (see text).}\label{TLH}
\end{figure}

The characteristic peaks observed under magnetic fields are mostly observed in dilute magnetic alloys~\cite{felsch1973magnetoresistivity} and strongly interacting quantum dots~\cite{PhysRevLett.81.5225}. A detailed NRG calculation has previously been done for the $\rho(T,B)$ of dilute magnetic alloy Ce$_x$La$_{1-x}$Al$_2$, $x = 0.0063$~\cite{Costi2000kondo}. Comparing with the NRG calculation, our modified Hamann expression demonstrates a valid extension of $\rho(T,B)$ to $T \ll T_{\rm{K}}$. It is worth mentioning that, in heavy fermion Ce$_x$La$_{1-x}$Cu$_6$, a continuous crossover from the single impurity Kondo system to the coherent Kondo lattice occurs with increasing the Ce impurity concentration~\cite{ONUKI1987281}. The resistivity for the concentrated Ce samples ($x >$ 0.5) has maximum below $T_{\rm{K}}$, which drops steeply with decreasing temperature. This type of maximum in $\rho(T)$ under zero magnetic field, observed in Kondo lattice materials, is due to the RKKY interactions between a periodic arrangement of Kondo ions.

To systematically study the characteristic peaks in $\rho(T,B)$ of 1$T$-VTe$_2$ single crystals, further investigation of the magnetic-field-induced splitting of Kondo resonance is provided by the MR measurements performed under very high magnetic fields up to 35 T (the magnetic field is applied perpendicular to the $c$-axis).  At 1.3 K, the negative MR reaches -5.2\% at 35 T. Figure~\ref{TLH} shows the resistivity versus $T^2$ measured under different magnetic fields. As can be seen, the peak in $\rho(T, B)$ is clearly visible at lower fields, but flattens out with increasing magnetic fields. For magnetic fields below 23 T, the data were analyzed by using the modified Hamann expression Eq. (6) and (8). At high magnetic fields, the anomaly is no longer detectable, and the temperature dependence of the resistivity above 23 T could be fitted to $\rho(T) = \rho_0 + \rho_{\rm{K}} + aT^2$. It is clear that the negative MR does not saturate after the anomaly is suppressed. This might indicate that higher magnetic fields are required to fully suppress the Kondo effect. The fitting parameters are included in Table I. It is worth noting that the Fermi liquid behavior (the $a$ coefficient, in particular) at high temperatures $T > T_{\rm{K}}$ and zero magnetic field is roughly the same as the Fermi liquid slope at low temperatures and 35 T (see Table I). As shown by the NRG calculations~\cite{Costi2000kondo}, the magnetic field splits and suppress the Kondo resonance which is due to the interaction between the magnetic impurity and the conduction electrons. Our results demonstrate that the Kondo effect is gradually suppressed and the system shows a crossover to the Fermi liquid state under high magnetic fields.

\begin{table} 
\centering
\caption{Parameters obtained from analysis of $\rho(B, T)$ in Fig.~\ref{MR} and Fig.~\ref{TLH} where $S$ =1/2 and $T_{\rm{K}}$ = 12 K are fixed.}
\begin{tabular}[t]{cccccc}
 \hline \hline
 $B$ (T) & $\rho_0 + \rho_{\rm{K}}$ ($\mu\Omega$ cm) & $a$ & $\rho_{\rm{K}}$ ($\mu\Omega$ cm) & $T_{\rm{W}}$ (K) & $\mu$ ($\rm{\mu_B}$)\\ 
 \hline
 $B \parallel c$ &   &  &  &  & \\
 0 & 77.1371 & 0.00455 & 7.2712 & 4.897   & \\ 
 3 & 77.0111 & 0.0045   & 6.9683 & 3.4324  & 2.4362 \\
 5 & 76.9279 & 0.00449 & 6.7993 & 3.8448 & 1.8888 \\
 7 & 76.8132 & 0.00447 & 6.5191 & 4.4409 & 1.6457 \\
 9 & 76.6433 & 0.00439 & 5.9707 & 4.923   & 1.4798 \\
\hline
 $B \perp c$ & &  &  &  & \\
0      &76.497	&0.005	&6.3749	&4.4231	& \\ 
2	&76.5056	&0.00488	&6.1447	&3.7431	&2.0637 \\ 
4	&76.469	&0.00467	&5.5658	&2.9452	&1.3954 \\ 
8	&76.4011	&0.00457	&5.3078	&3.9616	&1.0938 \\  
12	&76.1831	&0.00439	&4.2895	&4.1695	&0.8927 \\ 
16	&76.0906	&0.00464	&4.7916	&6.4127	&0.9462 \\ 
20	&75.6252	&0.00406	&2.2738	&5.3276	&0.7311 \\ 	
23	&75.3252	&0.00337	& & & \\
30	&74.8073	&0.00378	& & & \\  					
35	&74.5179	&0.00396	& & & \\ 		
 \hline \hline
\end{tabular}
\end{table}

\subsection{Hall effect}

\begin{figure}[htbp]
\includegraphics[width=0.5\textwidth]{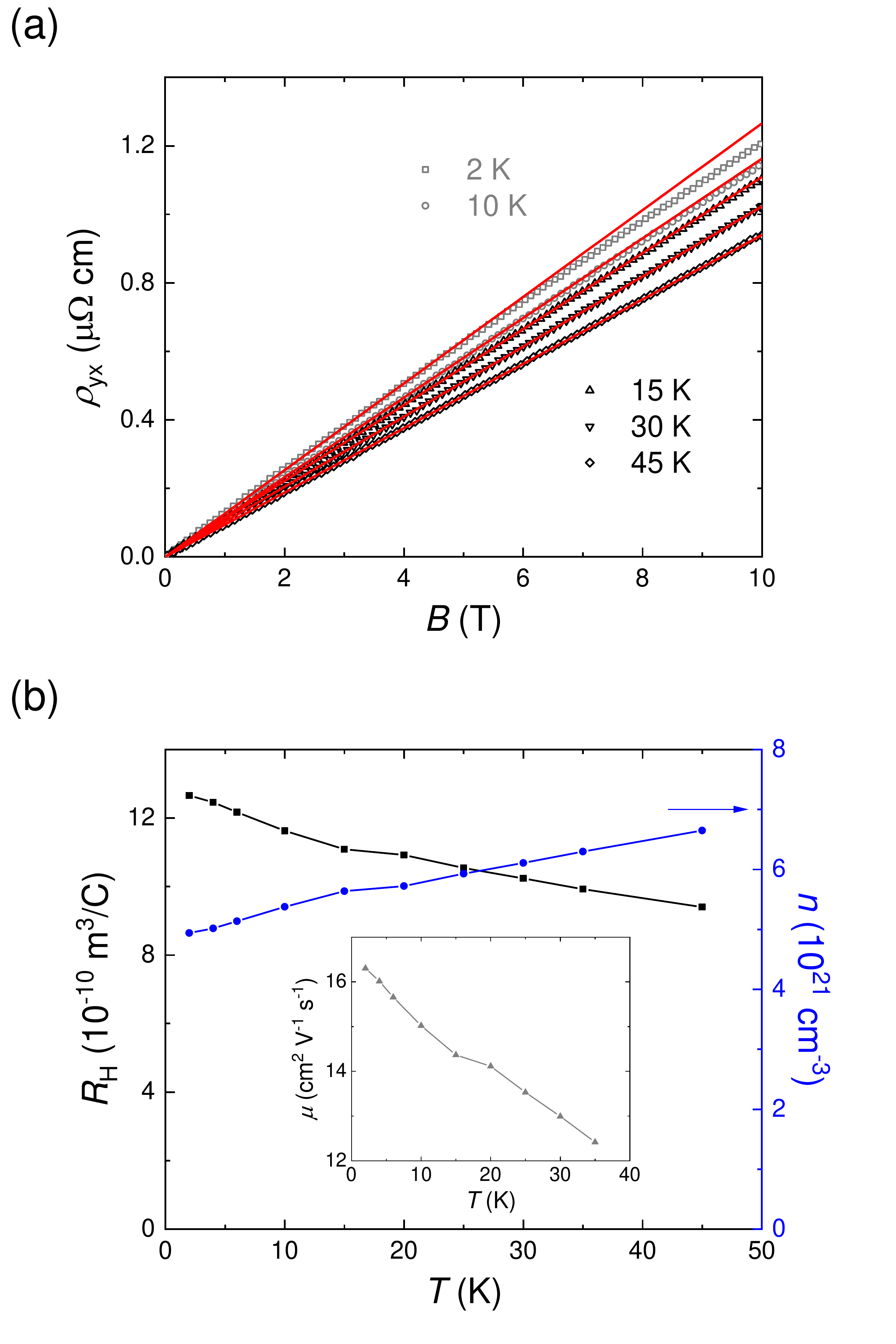}
\caption {(color online) (a) Magnetic field dependence of Hall resistivity at various temperatures of VTe$_2$. Red lines are linear fit from 0 to 5 T. (b) Temperature dependence of the Hall coefficient and effective charge carrier density. Inset: Hall mobility $\mu_{\rm{H}}$ vs temperature at $B$ = 5 T.}\label{Hall}
\end{figure}

Figure~\ref{Hall}(a) shows the magnetic field dependence of the Hall resistivity $\rho_{\rm{yx}}(B)$ measured at different temperatures up to 10 T. The red line is a linear fit of $\rho_{\rm{yx}}(B)$ from 0 to 5 T, characteristic of the ordinary Hall effect. By having a scrutiny to $\rho_{\rm{yx}}(B)$, the deviation from the linearity occurs at high magnetic fields below $T_{\rm{K}}$, which is consistent with the splitting of the Kondo resonance by the magnetic field. The Hall coefficient, $R_{\rm{H}} = \rho_{\rm{yx}}/B$ at 5 T, is plotted versus temperature in Fig.~\ref{Hall}(b). The positive $R_{\rm{H}}$ over the whole temperature range reveals that the electrical conduction is dominated by hole-type charge carriers. In contrast to the electron-type carriers in the VTe$_2$ nanoplates~\cite{VTe2Nano}, our results are consistent with the hole-like bands in the monolayer 1$T$-VTe$_2$~\cite{ARPES} and the result of recent angle-resolved photoemission spectroscopy measurements of 1$T$-V$_{1-x}$Ti$_x$Te$_2$, which characterize the circular and triangular hole-type Fermi surfaces around $\Gamma$ and K points, respectively~\cite{Mitsuishi2020}. As shown in Fig.~\ref{Hall}(b), the $R_{\rm{H}}$ exhibits a temperature dependency that differs from the expected one band approximation in simple metals. This might indicate a complex electronic structure in VTe$_2$, where the bands have temperature-dependent carrier concentrations and mobilities. Furthermore, $R_{\rm{H}}$ increases with decreasing temperature with an anomaly at $T \sim T_{\rm{K}}$, which is characteristic for the Kondo effect~\cite{Costi_1994}. The blue curve displays the temperature dependence of the effective charge-carrier density $n$, obtained by the formula $R_{\rm{H}} = 1/ne$ (one band approximation). At 45 K, the estimated concentration of free electrons is $6.7 \times10^{21}$ cm$^{-3}$. This value of $n$, due to the crude approximation that neglects the semimetallic character of the compound, can be considered as an upper limit of the real concentration in this material. The temperature dependence of the Hall mobility $\mu_{\rm{H}} (T)$ calculated with the single-band scenario at 5 T is shown in the inset of Fig.~\ref{Hall}(b). At 35 K, the value of $\mu_{\rm{H}}$ is 12.42 cm$^2$ V$^{-1}$ s$^{-1}$.

\section{CONCLUSIONS}

In summary, we have successfully synthesized single crystals of 1$T$-VTe$_2$ and measured their detailed magnetic and magnetotransport properties at low temperatures and under high magnetic fields. The magnetic susceptibility, electrical resistivity, and Hall effect are systematically studied revealing the presence of the Kondo behavior in this material. From the negative MR measurements, we point out that the intercalated V may act as the Kondo impurities. The observation of a peak below $T_{\rm{K}}$ in $\rho(T,B)$ reflects the splitting of the Kondo resonance. The analysis by using the modified Hamann expression is in good agreement with the experimental results. Furthermore, we directly show that the Kondo behavior is suppressed by strong magnetic fields, and for the fields above 23 T the bulk VTe$_2$ material behaves as an ordinary Fermi liquid system.

\section{acknowledgments}
This work is supported by the National Natural Science Foundation of China (Grant No. A0402/11534005 and A0402/11674164). X.D. acknowledges support from INLs LDRD program (19P43-013FP). K.G. acknowledges support from the US DOE BES Energy Frontier Research Centre "Thermal Energy Transport under Irradiation" (TETI). L.B. is supported by DOE-BES through award DE-SC0002613. The National High Magnetic Field Laboratory is supported by the National Science Foundation Cooperative Agreement No. DMR-1644779 and the State of Florida.

\bibliography{VTe2}

\end{document}